\documentclass[12pt]{article}
\usepackage{amssymb,epsf,graphicx}
\newcommand{\be}{\begin{equation}}
\newcommand{\ee}{\end{equation}}
\newcommand{\ben}{\begin{eqnarray}\displaystyle}
\newcommand{\een}{\end{eqnarray}}
\newcommand{\sectiono}[1]{\section{#1}\setcounter{equation}{0}}

\newcommand\crbig{\\\noalign{\vspace {1.5mm}}}
\newcommand\crsmall{\\\nopagebreak{\vspace {-6pt}}}
\newcommand{\csqrt}{\sqrt[\mp]}
\newcommand{\sgn}{\rm sgn}

\def\Im{\mathop{\rm Im}}
\def\Re{\mathop{\rm Re}}

\def\A{{\cal A}}
\def\B{{\cal B}}
\def\V{{\cal V}}

\begin{document}
{}~ \hfill\vbox{\hbox{hep-th/0408067}\hbox{KCL-MTH-04-11}}\break
\vskip 2.1cm

\centerline{\large \bf Closed Bosonic String Field Theory At Quartic Order}
\vspace*{8.0ex}

\centerline{\large \rm Nicolas Moeller}

\vspace*{8.0ex}

\centerline{\large \it Department of Mathematics}
\centerline{\large \it King's College London,}
\centerline{\large \it London, WC2R 2LS, UK} \vspace*{2.0ex}
\centerline{E-mail: {\tt moeller@mth.kcl.ac.uk}}

\vspace*{6.0ex}

\centerline{\bf Abstract}
\bigskip
We give a complete numerical description of the geometry of the
four-point contact interaction of closed bosonic string field
theory. Namely, we compute the boundary of the relevant region of the
moduli space of the four-punctured spheres, and everywhere in this
region we give the local coordinates around each punctures in terms of
a Strebel quadratic differential and mapping radii. The numerical
methods are explained in details. And the results are translated into
fits, which can in principle be used to compute the contact
interaction of any four off-shell string states.

\vfill \eject

\baselineskip=16pt

\tableofcontents

%%%%%%%%%%%%%%%%%%%%%%%%%%%%%%%%%%%%%%%%%%%%%%%%%%%%%%%%%%%%%%%%%%%%%%%%%%%%%%
\sectiono{Introduction}
\label{s_intro}
%%%%%%%%%%%%%%%%%%%%%%%%%%%%%%%%%%%%%%%%%%%%%%%%%%%%%%%%%%%%%%%%%%%%%%%%%%%%%%

Although a field theory of closed bosonic strings is known
(\cite{CSFT, Saad-Zwie, area}), until now only very little progress
has been made in understanding the fate of the bulk tachyon of the
closed bosonic string theory. Yet, calculating the tachyon potential,
and rewriting the theory around one of its minimum, is probably the
first thing that one should do if one is given a field theory with a
tachyonic instability. The problem with closed bosonic string field
theory (CSFT) is its complexity; unlike open string field theory,
which is cubic in the string field, CSFT is non-polynomial. This adds
a great deal of complexity, because all terms higher than cubic will
involve an integration over a subset of the moduli space of the
relevant Riemann surfaces; and at every point of this restricted
moduli space we need to specify local coordinates, so that we can
properly insert the off-shell vertex operators.

\paragraph{}
It may be nevertheless feasible to do calculations in CSFT, by
truncating the action to some finite polynomial order; it is possible
that the results computed from the truncated action converge as the
truncation order is increased.  If we can do so, one of the most
important questions that we would like to answer, is whether the
tachyon potential has a minimum or not. If it has a global minimum,
then it is natural to expect that the theory expanded around this
minimum may be a superstring theory. The mechanism by which the
bosonic theory would flow to a superstring theory by tachyon
condensation, would have to involve spontaneous breaking of the
twenty-six dimensional Poincar\'e symmetry. Evidence that this can
happen in CSFT at low order, was given in \cite{West}. This is also
supported by the fact that all the ten-dimensional superstring
theories can be realized from the closed bosonic string theory after
some suitable compactification and truncation of the spectrum
(\cite{fermionic}). But although the tachyon potential has a minimum
to third order in the string field (as was originally found in
\cite{Kost-Samu}), to quartic order it doesn't have a minimum anymore,
as was shown in \cite{Belo}. Clearly, to get some new insight into
this question, we need to express the tachyon potential to higher
orders.

Another interesting question, which has appeared more recently in the
literature (see \cite{orbifolds}, \cite{Okaw-Zwie1} and references
therein), is that of the condensation of closed tachyons localized on
the apex of an orbifold. It has been conjectured that the condensation
of such tachyons changes the background geometry to that of an
orbifold with a lesser deficit angle, until it eventually condenses to
flat space. A numerical check of this conjecture was attempted in
\cite{Okaw-Zwie1} in closed bosonic string theory. At low order and
truncation level, a relatively good agreement (of roughly 70\%) was
found.  Here too, a more thorough study of the problem is very likely
to require contact interactions of higher orders.

\paragraph{}
In \cite{Belo}, Belopolsky calculated the four-tachyon contact term.
His method relied on analytical properties of the quadratic
differentials on a four-punctured sphere (although numerical
computations were still needed in order to extract the amplitude).
Unfortunately, these analytical methods don't seem to generalize
obviously to the contact interactions of more than four strings.  The
goal of the present paper is to provide an efficient numerical method
for calculating the geometry of the contact interactions, that does
generalize to higher order interactions. We also give our numerical
results, that describe completely the geometry of the interaction
everywhere in the moduli space. They can thus be used to calculate the
contact amplitude of any four off-shell string states.

\paragraph{}
The structure of the paper is as follows: First, in Section
\ref{s_geometry}, we briefly review how to calculate the contact
amplitudes with the help of quadratic differentials, and how the local
coordinates at the punctures can be computed from the Strebel
quadratic differential. We review the moduli space of the contact
interaction in Section \ref{s_moduli}. In section \ref{s_a}, we expose
in details our numerical method to find the Strebel differential at a
given point in the moduli space; then we go on and describe, in
Section \ref{s_rho}, our numerical method to compute the mapping radii
of the Strebel differential. We then present our numerical results in
Section \ref{s_results}, in the form of fits. And in Section
\ref{s_4tachyon}, we show that the result of \cite{Belo} for the
four-tachyon contact term is confirmed by our results. Finally,
Section \ref{s_conclusions} is devoted to discussions and conclusions.

%%%%%%%%%%%%%%%%%%%%%%%%%%%%%%%%%%%%%%%%%%%%%%%%%%%%%%%%%%%%%%%%%%%%%%%%%%%%%%
\sectiono{The geometry of the contact interaction}
\label{s_geometry}
%%%%%%%%%%%%%%%%%%%%%%%%%%%%%%%%%%%%%%%%%%%%%%%%%%%%%%%%%%%%%%%%%%%%%%%%%%%%%%

In this section, we summarize very briefly the notions that are
necessary to compute off-shell contact amplitudes. More detailed
explanations can be found in \cite{Saad-Zwie, Belo, Belo-Zwie,
Strebel}.

\paragraph{}
To calculate an off-shell contact amplitude of $N$ string states on an
$N$-punctured sphere, one must insert the vertex operators at the
punctures, transforming them according to the local coordinates at the
punctures. One then calculates their correlation function after
inserting antighosts. At last, one has to integrate the correlators on
the relevant subset\footnote{In general $\V_{g,N}$ is the restricted
moduli space of the $N$-punctured Riemann surfaces of genus
$g$. Although we will consider here only the case $g=0$, we will keep
this notation.}  $\V_{0,N}$ of the moduli space of $N$-punctured
spheres.  The geometry of the punctured sphere that determines the
local coordinates at the punctures, is given by the metric of minimal
area\footnote{ Actually, since in our case the area is infinite, we
have to minimize the {\it reduced} area. See \cite{area} for more
details.}  with the constraint that every nontrivial closed curve has
length $\geq 2 \pi$.  It follows that the geometry of the punctured
sphere is that of $N$ semi-infinite cylinders of circumference $2
\pi$, joining along the edges of a polyhedron whose vertices always
join three edges. The constraint on the nontrivial closed curves
ensures that we do not over-count Feynman diagrams. The solution to
this minimal area metric is given by a Strebel quadratic differential,
which we now briefly introduce.

\paragraph{}
A quadratic differential $\varphi$ is a geometrical object that lives
on a Riemann surface. In the local coordinate $z$, it takes the form
$\varphi = \phi(z) (dz)^2$; and it is invariant under conformal
changes of local coordinates, in the sense that, if $w$ is another
coordinate, we have \be \phi(z) (dz)^2 = \phi(w) (dw)^2
\,. \label{coordchange} \ee A quadratic differential defines a metric,
whose length element is $ds = \sqrt{|\phi(z)|} \, |dz|$. In
particular, in a neighborhood of a second order pole $z = p$, \be
\phi(z) = {r \over (z-p)^2} + {\cal O}\left((z-p)^{-1}\right) \,,
\label{secpole} \ee and if $r < 0$, the geometry is that of a
semi-infinite cylinder of circumference $-2 \pi r$. The quadratic
differentials that we will consider here are regular everywhere except
at the punctures, where they have second order poles.  We will call
the quantity $r$ in (\ref{secpole}), the {\it residue} of $\varphi$ at
the puncture $p$. It is easy to see from (\ref{coordchange}), that the
notions of pole and of residue, as well as the zeros of $\varphi$, are
conformal invariants. If $\varphi$ has a second order pole with
residue $r$ at $z = p$, then there is a local coordinate $w$, defined
in a neighborhood of the puncture, in which \be \varphi = {r \over
w^2} (dw)^2 \,. \label{natural} \ee We will call such a local
coordinate $w$, a {\it natural coordinate} at the puncture $z = p$.

A {\it horizontal trajectory} of $\varphi$ is a maximal curve on which
$\phi(z) (dz)^2 > 0$, where $dz$ is tangential to the curve. For
example, the horizontal trajectories of (\ref{natural}) are circles
centered at the puncture $w = 0$.  A {\it critical} horizontal
trajectory is a horizontal trajectory one of whose ends is either a
zero or a pole of order one. We note here that, as one can easily
verify, a zero of order $n$ is the source of $n+2$ horizontal
trajectories.  The {\it critical graph} of $\varphi$ is the set formed
by all critical horizontal trajectories.

\begin{figure}[!ht]
\begin{center}
\input{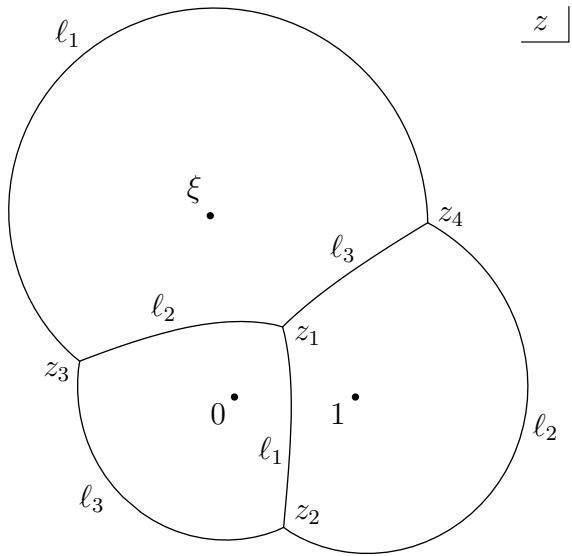}
\caption{\footnotesize{The critical graph of the Strebel differential,
whose poles are at $0$, $1$, $\xi$, and $\infty$; here we have chosen
the particular value $\xi = -0.2 + 1.5 \, i$. It has the topology of a
tetrahedron, with a puncture on each side. Its zeros are labelled
$z_i$, $i = 1,..,4$. We have also labelled the edge lengths $\ell_i$
(note that two opposite edges have the same length).}}
\label{crit_graph_f}
\end{center}
\end{figure}

\paragraph{}
From now on, we will only consider the case of the $4$-punctured
spheres, which we will describe in the uniformizer coordinate $z$.
The last definition\footnote{ This definition can be made much more
general, but for simplicity, we restrict ourselves here to the
four-punctured spheres.} that we need is that of the {\it Strebel
differential}.  It is a quadratic differential whose four poles are
second order poles with residue $-1$, and whose critical graph has the
topology of a tetrahedron. In particular, the Strebel differential has
four simple zeros, which are the vertices of the tetrahedron, at which
three critical trajectories (edges) are joining.  See figure
\ref{crit_graph_f} for the critical graph of a particular Strebel
differential.
On this figure we have also labelled the lengths (in the
$\varphi$-metric) of the edges of the tetrahedron. From the residue
conditions, it is easy to see that two opposite edges (which don't
touch each other) have the same length. Also we have that the
circumference of each face is $2 \pi$, in particular $\ell_1 + \ell_2
+ \ell_3 = 2 \pi$, and thus we have only two independent lengths, for
example $\ell_1$ and $\ell_2$.

Using the ${\rm PSL}(2, \mathbb{C})$ automorphism 
group of the Riemann sphere, we can always place three punctures at, 
respectively, $z = 0$, $1$, and $\infty$. The fourth puncture will be 
called $\xi$. The residue condition for the Strebel differential, 
is satisfied by the general form 
\ben
\phi(z) &=& {-(z^2 - \xi)^2 \over z^2 \, (z-1)^2 \, (z-\xi)^2} + 
{a \over z \, (z-1) \, (z-\xi)} \nonumber \crbig
&=& {-z^4 + a \, z^3 + (2 \xi - (1+\xi)a) \, z^2 + a \xi \, z - \xi^2 \over 
z^2 \, (z-1)^2 \, (z-\xi)^2} \,, \label{phi1}
\een
where $a$ is any complex number. The puncture conditions at $0$, $1$ and $\xi$
are straightforwardly verified from the above form. For the puncture
at infinity, we need to change the coordinate to $w = 1/z$, and we should
verify that $\phi(w) = -1/w^2 + {\cal O}(w^{-1})$. This is easily done:
\ben
&& \phi(z) (dz)^2 = \phi(w) (dw)^2 \nonumber \\
& \Rightarrow & \phi(w) = \phi(z = 1/w) \left({dz \over dw}\right)^2 
= \phi(z = 1/w) \, {1 \over w^4} = -{1 \over w^2} + {\cal O}(w^{-1}) \,, \nonumber
\een
where by $\phi(z = 1/w)$ we mean the function $\phi(z)$ in the
coordinate $z$, evaluated at $z = 1/w$. And we see that (\ref{phi1})
is therefore the most general form satisfying the residue condition.

The metric defined from the Strebel differential is a metric of minimal 
area. Moreover, by a theorem of Strebel (\cite{Strebel}), 
the Strebel differential for a given $\xi$ is unique. Therefore, if all 
nontrivial closed curves on the punctured sphere, have length $\geq 2 \pi$, 
we will have solved our minimal area problem.
One of the results of this 
paper is to determine numerically the coefficient $a$ in (\ref{phi1}), and 
therefore $\varphi$, for all values of $\xi$ in the restricted moduli space.

\begin{figure}[!ht]
\begin{center}
\input{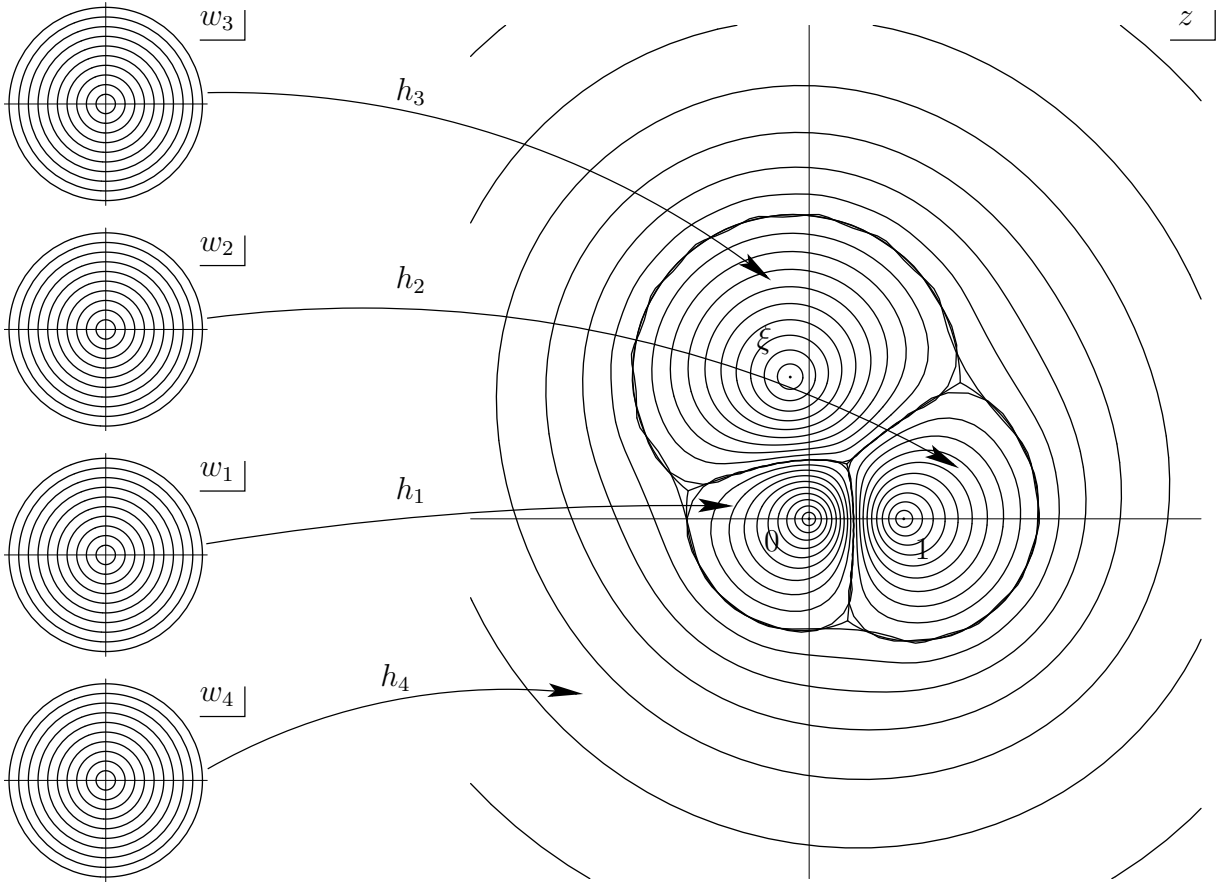}
\caption{\footnotesize{The four conformal maps as series (\ref{conf2}) to order $18$, 
are represented for the particular 
value $\xi = -0.2 + 1.5 \, i$ by mapping circles of radii $k/{10}$ for $k = 1, .. ,10$. Also plotted 
is the critical graph of the quadratic differential.}}
\label{conformal_f}
\end{center}
\end{figure}

\paragraph{}
The Strebel differential defines (up to unimportant phases) the 
conformal maps from the four 
punctured disks (with local natural coordinates $w_n$) 
to a $4$-punctured sphere, in such a way that the punctures 
$w_n = 0$ are mapped to the punctures $z = p_n$, and the boundaries $|w_n| = 1$ of 
the unit disks are mapped to the part of the critical graph surrounding $p_n$ (see 
Figure \ref{conformal_f}). Our notation is that the punctures are labelled by the 
index $n$ running from $1$ to $4$, with $p_1 = 0$, $p_2 = 1$, $p_3 = \xi$ and 
$p_4 = \infty$. The method for expressing explicitly these conformal 
maps from the knowledge of $\varphi$ was described in
\cite{Belo-Zwie}. First, we expand $\varphi$ in 
a Laurent series around the puncture $p_n$. 
\be
\varphi = \left( -{1 \over (z-p_n)^2} + {b_{-1} \over z-p_n} + b_0 + b_1 (z-p_n) 
+ \ldots \right) (dz)^2 \,. \label{conf1}
\ee
The coefficients $b_i$ are calculated straightforwardly from (\ref{phi1}).
The map from the punctured unit disk to the punctured sphere, is 
expanded as follows\footnote{Our notation here differs from \cite{Belo-Zwie}. 
It is convenient to expand in powers of $(\rho_n w_n)$, instead of $w_n$.}
\be
z = h_n(w_n) = p_n + (\rho_n w_n) + d_1 \, (\rho_n w_n)^2 + d_2 \, (\rho_n w_n)^3 + \ldots \,.
\label{conf2}
\ee
Here $\rho_n = \left| {d z \over d w_n} \right|_{w_n = 0}$ is the {\it mapping radius} 
of the conformal map $h_n$.
In the natural coordinates $w_n$, the quadratic differential has the form
\be
\varphi = -{1 \over w_n^2} (dw_n)^2 \label{conf3} \,.
\ee
Now we use (\ref{conf1}), (\ref{conf2}) and (\ref{conf3}) to calculate recursively 
the coefficients $d_i$ from the $b_i$'s. For the first few coefficients, we get
\ben
d_1 &=& {1 \over 2} \, b_{-1} \nonumber \\ 
d_2 &=& {1 \over 16} \left(7 \, {b_{-1}}^2 + 4 \, b_0 \right) \nonumber \\
d_3 &=& {1 \over 48} \left(23 \, {b_{-1}}^3 + 28 \, b_{-1} b_0 + 8 \, b_1 \right) \nonumber \\
d_4 &=& {1 \over 768} \left(455 \, {b_{-1}}^4 + 856 \, {b_{-1}}^2 b_0 + 144 \, {b_0}^2 + 
368 \, b_{-1} b_1 + 96 \, b_2 \right) \,. \nonumber \\
&\vdots& \nonumber 
\een
Note that the mapping radii cannot be calculated from the above recursive method. 
But they can be calculated, for example, by the formula (\ref{rho_Bel_Zwi}) 
derived in \cite{Belo-Zwie}.
As a consistency check of our results, we have calculated the conformal maps 
as a series (\ref{conf2}) to order $18$, for 
the particular value $\xi = -0.2 + 1.5 \, i$. We find $a = 1.12245 + 1.23948 \,i$, 
and 
$$
\begin{array}{rcl}
\rho_1 &=& 0.69819 \\
\rho_2 &=& 0.88628 \\
\rho_3 &=& 1.34118 \\
\rho_4 &=& 0.46138 \nonumber \,.
\end{array}
$$
In figure \ref{conformal_f}, we have represented the four conformal maps by mapping circles of 
radii $k/{10}$ for $k = 1, .. ,10$.

%%%%%%%%%%%%%%%%%%%%%%%%%%%%%%%%%%%%%%%%%%%%%%%%%%%%%%%%%%%%%%%%%%%%%%%%%%%%%%
\sectiono{The moduli space}
\label{s_moduli}
%%%%%%%%%%%%%%%%%%%%%%%%%%%%%%%%%%%%%%%%%%%%%%%%%%%%%%%%%%%%%%%%%%%%%%%%%%%%%%

As already mentioned, a point $\xi$ belongs to the moduli space
$\V_{0,4}$ of the contact interaction (also called {\it restricted}
moduli space), only if all nontrivial closed curves on its
corresponding tetrahedron have length $\geq 2 \pi$. One can easily
check that this is equivalent to the conditions
\be 
\ell_1, \, \ell_2, \, \ell_3 \leq \pi \,,
\ee
where the $\ell_i$ are the edge lengths (see Figure \ref{crit_graph_f}). 
We show $\V_{0,4}$ in Figure \ref{moduli_f}. It is delimited by 
three curves $\B_1$, $\B_2$ and $\B_3$ on which, respectively 
$\ell_1$, $\ell_2$ and $\ell_3$ are equal to $\pi$.

\begin{figure}[!ht]
\begin{center}
\input{moduli.pstex_t}
\caption{\footnotesize{The moduli space $\V_{0,4}$. It is bounded by the
three curves $\B_1$, $\B_2$ and $\B_3$.  We need only to concentrate
on the space $\A$ (shaded in darker grey), bounded by $\B_1$, the unit
circle at the origin, and the line $\Re z = 1/2$; all the other
regions can be reached from $\A$ by ${\rm PSL}(2, \mathbb{C})$
transformations and complex conjugations.}}
\label{moduli_f}
\end{center}
\end{figure}

We don't need to calculate the Strebel differential everywhere 
on this space because there are some symmetries. First, there is the obvious 
symmetry $\xi \rightarrow \xi^*$. 
Then there are six ${\rm PSL}(2, \mathbb{C})$ transformations that 
permute the three fixed poles $\{0, 1, \infty\}$. They are generated by 
\ben
g_1:&\, & \xi \rightarrow 1-\xi \\
{\rm and} \qquad g_2:&\, & \xi \rightarrow 1/\xi \,.
\een
In total, $\V_{0,4}$ is thus made of twelve parts separated by the
real axis, the line $\Re(\xi) = 1/2$, and two circles of unit radius
centered on $0$ and on $1$ respectively. In our computations, we will
therefore only consider the subspace $\A$, which is bounded from the
right by $\{\xi : \ \Re(\xi) = 1/2\}$, from below by the unit circle
centered on $0$, and from the left and the top by the curve $\B_1$ (see
Figure \ref{moduli_f}).

To find the parameter $a$ in the other regions of $\V_{0,4}$, we apply, 
for ${\rm PSL}(2, \mathbb{C})$ transformations $\xi \rightarrow g(\xi)$,
the transformation law (\ref{coordchange}) to $\varphi$ given in (\ref{phi1}); 
and when $\xi \rightarrow \xi^*$, we have $a \rightarrow a^*$. 
The transformation law for the mapping radius $\rho_n$ at the puncture $p_n$, 
is 
\be
\rho_n \rightarrow \left|g'(p_n)\right| \, \rho_n \,. \label{rhotransf}
\ee
And it is invariant under complex conjugation $\xi \rightarrow \xi^*$. 
Note that if $g$ in (\ref{rhotransf}), sends the finite puncture $p_n$ 
to infinity, one must calculate its transformation in the coordinate 
$w = 1/z$. In particular, if $g(\xi) = 1/\xi$, the mapping radii at the 
punctures $z = 0$ and $z = \infty$ are invariant.

\paragraph{}
Note that we don't know the analytic expressions of the curves $\B_i$.
We thus have to calculate one of them, say $\B_1$, numerically. This
is done by finding the points where $\ell_1 = \pi$; the result will be
presented in Section \ref{s_results}. But we can already understand a
few points of this moduli space. First, when $\xi = 1/2$, we can show
that the Strebel differential is given by $a=2$, and then one can
verify that $\ell_2 = \ell_3 = \pi$.  This shows that at the point
$\xi = 1/2$, the two curves $\B_2$ and $\B_3$ intersect. And
therefore, from the symmetry of $\V_{0,4}$, one also has that, at
$\xi=-1$, $\B_1$ and $\B_2$ intersect, while at $\xi=2$, it is $\B_1$
and $\B_3$ that intersect. The knowledge that the point $\xi=-1$
belongs to $\B_1$, will be used in Section \ref{s_results}, when we
express $\B_1$ as a fit. Note that all these points on the real axis
are singular, in the sense that the Strebel differentials at those
points have double zeros, which means that the tetrahedron is
singular, two of its edges having collapsed to zero length.

The point $Q \equiv {1 \over 2} + {\sqrt{3} \over 2} \, i$ is also interesting. 
Indeed, when $\xi = Q$, the three punctures $0$, $1$ and $\xi$ are placed 
at the corners of an equilateral triangle. In fact, this is the most 
symmetric configuration, whose corresponding tetrahedron is a regular 
tetrahedron. One can check that for this configuration, 
$a = 2 + {2 \over \sqrt{3}} \, i$.

%%%%%%%%%%%%%%%%%%%%%%%%%%%%%%%%%%%%%%%%%%%%%%%%%%%%%%%%%%%%%%%%%%%%%%%%%%%%%%
\sectiono{Computation of the Strebel differential}
\label{s_a}
%%%%%%%%%%%%%%%%%%%%%%%%%%%%%%%%%%%%%%%%%%%%%%%%%%%%%%%%%%%%%%%%%%%%%%%%%%%%%%

Here we explain in details how to find the parameter $a$ in (\ref{phi1}) 
such that the quadratic differential is the Strebel differential 
for the specified position of the fourth puncture $\xi$. Once $a$ is found, 
the mapping radii can be computed relatively easily; this will be described 
in the next section. The numerical results for $a$ and the mapping radii will 
be presented in Section \ref{s_results}.

\paragraph{}
We define the {\it complex length} of a curve $C$ by $\ell(C) \equiv
\int_{C} \sqrt{\phi(z)} dz$ or, if $C$ is a segment from $z_i$ to $z_j$,
\be 
\ell(z_i, z_j) \equiv \int_{z_i}^{z_j} \sqrt{\phi(z)} dz \label{comp_length} \,.
\ee
The square root must be chosen such that its branch cut doesn't intersect 
the path of integration. Now suppose that $z_i$ and $z_j$ are zeros of $\varphi$, 
and that $\varphi$ is the Strebel differential. This means that there is a critical 
trajectory, that we call $C$, joining $z_i$ and $z_j$. From the definition of a 
horizontal trajectory, we have that $\sqrt{\phi(z)} dz$ is real on $C$, and therefore 
$\ell(C)$ is real (and is equal to the length of $C$ in the $\varphi$-metric). 
By deforming the path of integration, we then have that 
$\ell(z_i, z_j)$ is real. 
Indeed, even if $\varphi$ has a pole in between 
$C$ and the segment $[z_i, z_j]$, the residue of this pole would change the integral only 
by a real quantity, because the residues (in the usual sense) of $\sqrt{\phi(z)}$ are 
all purely imaginary.

Therefore, in order to find the Strebel differential, we will look for the quadratic 
differential $\varphi$ that satisfies the condition that $\ell(z_i, z_j)$ 
must be real for any two zeros $z_i$ and $z_j$. Since the quadratic differentials 
(\ref{phi1}) that we are considering, are already constrained to have residues $-1$ 
at the four punctures, we easily see that the above reality condition is satisfied 
if $\ell(z_i, z_j)$ is real for two independent segments; for example
\be
\left\{
\begin{array}{rcl}
\Im \ell(z_1, z_2) &=& 0 \crbig
\Im \ell(z_1, z_3) &=& 0 \,,
\end{array} \right. \label{system}
\ee
where $z_1$, $z_2$ and $z_3$ are three zeros of $\varphi$.
Solving these two real equations will determine the complex parameter $a$.

\paragraph{}
To solve the system of equations (\ref{system}), we use the Newton 
method in two dimensions. Namely, we have two real equations for the 
two real unknowns $\Re a$ and $\Im a$. 
In general, the Newton method for the system of $n$ equations of $n$ unknowns 
$\vec{f}(\vec{x}) = 0$ consists of starting with a $\vec{x}_0$ not too far from 
the solution, and repeating the iteration
$$
\vec{x}_{i+1} = \vec{x}_i - J(\vec{x}_i)^{-1} \vec{f}(\vec{x}_i)
$$
until $\vec{x}_i$ converges. Here 
$J(\vec{x}_i) = \left({\partial \vec{f} \over \partial \vec{x}} 
\right)_{\vec{x} = \vec{x}_i}$ is the Jacobian. In order to calculate 
it in our situation, we note that for a holomorphic function $f(z)$ we have
${\partial \Im f(z) \over \partial \Re z} = \Im f'(z)$, and  
${\partial \Im f(z) \over \partial \Im z} = \Re f'(z)$.
Our Jacobian is thus 
\be
J = \left( \begin{array}{cc} {\partial \Im \ell(z_1, z_2) \over 
\partial \Re a} & 
{\partial \Im \ell(z_1, z_2) \over \partial \Im a} \crbig
{\partial \Im \ell(z_1, z_3) \over \partial \Re a} & 
{\partial \Im \ell(z_1, z_3) \over \partial \Im a} \end{array} \right)
= \left( \begin{array}{cc} \Im {d \over da} \ell(z_1, z_2) & 
\Re {d \over da} \ell(z_1, z_2) \crbig
\Im {d \over da} \ell(z_1, z_3) & \Re {d \over da} \ell(z_1, z_3) 
\end{array} \right) \,,
\ee
and we will need therefore to evaluate ${d \over da} \ell(z_i, z_j)$ 
as well as $\ell(z_i, z_j)$.

But before we go on and solve the system of equations (\ref{system}), 
we see that there is a 
problem with the computation of the complex lengths (\ref{comp_length}), 
namely we need a square
root whose branch cut we shall never intersect, although we have to
evaluate it along various paths whose positions we don't know {\it \`a
priori}. The solution to this problem, a square root without a branch cut,
is a central tool of our numerical methods; it is presented in 
the next subsection.

%%%%%%%%%%%%%%%%%%%%%%%%%%%%%%%%%%%%%%%%%%%%%%%%%%%%%%%%%%%%%%%%%%%%%%%%%%%%%%
\subsection{A square root without a branch cut}
%%%%%%%%%%%%%%%%%%%%%%%%%%%%%%%%%%%%%%%%%%%%%%%%%%%%%%%%%%%%%%%%%%%%%%%%%%%%%%
 
First we note that in all our calculations that present a potential
branch cut problem, the square root is to be evaluated inside an
integral along some path. It has therefore to be evaluated at points
that are "close to each other". It is thus quite natural to define a
{\it continuous square root} (that we denote by $\csqrt{\ }$) along
the path, in the following way: It has to remember the last result
$\csqrt{z_{n-1}}$ that it calculated; then to calculate the next
value, it computes the usual square root $\sqrt{z_n}$, and chooses its
sign $\csqrt{z_n} = \pm \sqrt{z_n}$ so that $|\csqrt{z_n} -
\csqrt{z_{n-1}}|$ is minimal.

\begin{figure}[!ht]
\begin{center}
\input{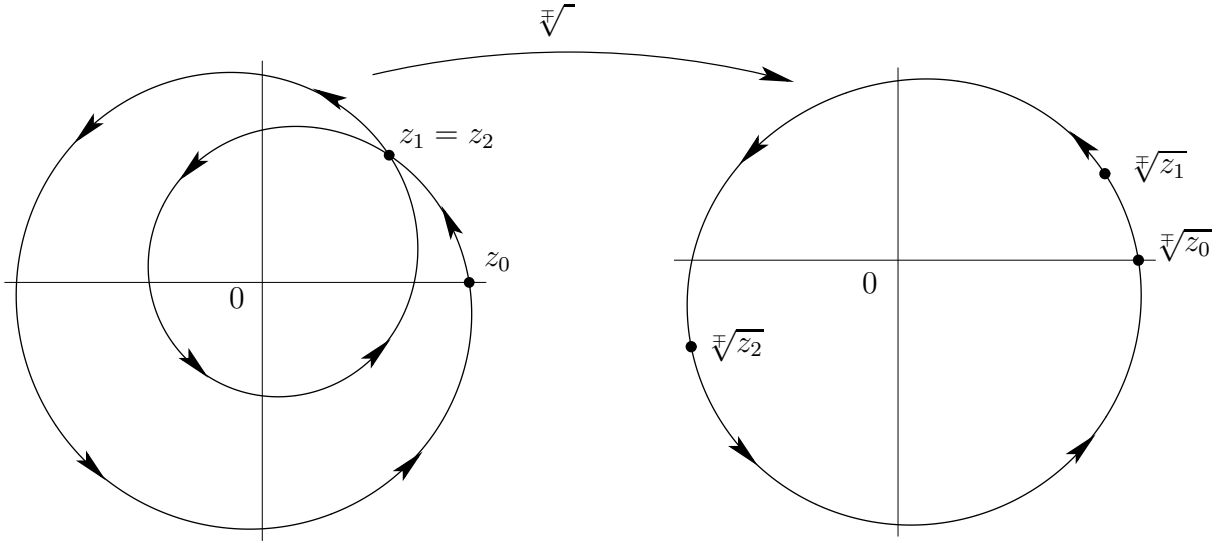}
\caption{\footnotesize{The image by the continuous square root of a
continuous closed path winding twice around the origin is a continuous
closed path winding once. Note that the point $z_1 = z_2$ that is met
twice by the path, is mapped to two different values.}}
\label{sqrt_f}
\end{center}
\end{figure}

We illustrate the result of evaluating the continuous square root along 
a path in Figure \ref{sqrt_f}, where the path is a closed
curve winding twice around the origin; its image is a closed path that
winds once around the origin. We see that the gain of avoiding the
branch cut (and therefore defining a continuous function), is at the
expense of the sign of the function not being globally
defined\footnote{The domain of $\csqrt{\ }$ should really be a double
cover of $\mathbb{C} - \{0\}$; it would then be a well-defined holomorphic
function.}. Indeed, in our example we have $\csqrt{z_1} \neq \csqrt{z_2}$ 
although $z_1 = z_2$. 
Practically, the continuous square root is implemented by the following 
(schematized) code: \\ \\ \nopagebreak
{\footnotesize 
complex $\csqrt{\ }\,$(complex $z$) \crsmall
\{ \crsmall
\hspace{0.5in} {\bf static} complex $p = 0$; \crsmall
\hspace{0.5in} complex $w$; \crsmall
\crsmall
\hspace{0.5in} $w = \sqrt{z}$; \crsmall
\hspace{0.5in} {\bf if} ($\Re(w) \Re(p) + \Im(w) \Im(p) < 0$) \crsmall
\hspace{1in} $w = -w$; \crsmall
\hspace{0.5in} $p = w$; \crsmall
\hspace{0.5in} {\bf return} $w$; \crsmall 
\}} \\ 
\\ 
The variable $p$ is declared ``static'', which means that its value is
retained between two calls of the function. The square root $w$ is
compared with the previous result $p$. If their ``scalar product''
$\Re(w) \Re(p) + \Im(w) \Im(p)$ is negative, we suspect that the
branch cut was crossed, and thus we flip the sign of $w$.

We emphasize that the continuous square root makes sense only when it
is evaluated along a path that does not meet the origin (practically,
it has to be evaluated at successive values that are close to each
other compared to their distance to the origin). We will thus 
only use it inside an integral
and won't bother about the sign in front of the integral. Luckily, for
the problems we need to solve, like (\ref{system}), we will always be
able to go around the fact that the global sign is ambiguous.

Finally, if the continuous square root is evaluated along a
non-self-intersecting path that does not go through the origin, we can
always draw a branch cut that does not meet the path. The holomorphic
square root with this given branch cut is then equal (up to a global
sign) to the continuous square root. The latter is therefore
holomorphic in some neighborhood of the path.

%%%%%%%%%%%%%%%%%%%%%%%%%%%%%%%%%%%%%%%%%%%%%%%%%%%%%%%%%%%%%%%%%%%%%%%%%%%%%%
\subsection{Numerical evaluation of $\ell(z_i, z_j)$}
%%%%%%%%%%%%%%%%%%%%%%%%%%%%%%%%%%%%%%%%%%%%%%%%%%%%%%%%%%%%%%%%%%%%%%%%%%%%%%

Without loss of generality, we take $z_i = z_1$ and $z_j = z_2$.
We need to numerically\footnote{Putting aside the branch cut problem, the 
integral (\ref{integral}) can actually be expressed in terms of elliptic 
functions. We still compute it numerically for two reasons. First, the analytic 
expression is quite complicated, and since we have to evaluate it numerically anyway 
in order to solve (\ref{system}), we wouldn't gain much. Second, (\ref{integral}) is 
integrable analytically only because the order of the polynomial in the square root 
is not greater than four (see for example \cite{elliptic}); and therefore an analytic 
evaluation of (\ref{integral}) would not generalize to contact interactions of order 
higher than four.} 
evaluate 
\be
\ell(z_1, z_2) = \int_{z_1}^{z_2} \csqrt{\phi(z)} dz = \int_{z_1}^{z_2}{
{\csqrt{-(z-z_1) (z-z_2) (z-z_3) (z-z_4)} \over z (z-1) (z-\xi)} dz} \label{integral}
\ee
where $z_1, \ldots , z_4$ are the four zeros of $\phi(z)$, and the integration is 
along a straight path. Parameterizing
\be
z(t) = {z_2+z_1 \over 2} + {z_2-z_1 \over 2} \, t = 
z_1 + {z_2-z_1 \over 2} \, (t+1) \,, \label{z(t)}
\ee
we can rewrite this integral as 
\be
\ell(z_1, z_2) = \left({z_2 - z_1 \over 2} \right)^2 \int_{-1}^1{{\csqrt{(z(t)-z_3) 
(z(t) - z_4)} \over z(t) (z(t)-1) (z(t)-\xi)} \sqrt{1-t^2} dt} \,, \label{int1}
\ee
where we have written the usual square root $\sqrt{1-t^2}$; this is fine since 
$(1-t^2)$ is never negative on the integration path. But
it is important to note here that we have made an arbitrary choice of sign in 
(\ref{int1}); indeed, since the global sign of the continuous square root is 
ambiguous, there is no way to tell which is the right sign in the identity 
$\csqrt{x y} = \pm \csqrt{x} \sqrt{y}$. But since we don't care about the global 
sign of $\ell(z_1, z_2)$, what sign we are choosing is not 
important, as long as we do the same choice of sign when computing 
${d \over da} \ell(z_i, z_j)$ (see (\ref{derivative}) and comment thereafter).

We want to evaluate the integral (\ref{int1}) numerically. But we observe 
that, in general, one or more poles will be close enough to the path of 
integration to
destroy the accuracy of the numerical integration method. To remedy this, we 
subtract the finite poles from the integrand and add back the result of
their analytic integrations:
\ben
\ell(z_1, z_2) &=& \left({z_2 - z_1 \over 2} \right)^2 \left\{ \int_{-1}^1{\left( 
{\csqrt{(z(t)-z_3) (z(t) - z_4)} \over z(t) (z(t)-1) (z(t)-\xi)}
- \sum_{n=1}^3{s_n r_n \over z(t) - p_n} \right) \sqrt{1-t^2} dt} \right. 
\nonumber \\
&& \hspace{75pt} \left. + \sum_{n=1}^3{s_n r_n \int_{-1}^1{
{1 \over z(t) - p_n} \sqrt{1-t^2} dt}} \right\} \,,
\label{length}
\een
where $p_1=0, p_2=1$ and $p_3=\xi$ are the finite poles, and 
\be
r_n = \sqrt{(p_n-z_3) (p_n - z_4)} \, \prod_{{m=1}\atop{m \neq n}}^3{(p_n-p_m)^{-1}} \,.
\ee
The signs $s_n = \pm 1$ are chosen in such a way that 
$$
\max_{t \in \left[-1,1\right]} \left\{ \left| {\csqrt{(z(t)-z_3) (z(t) - z_4)} 
\over z(t) (z(t)-1) (z(t)-\xi)} - \sum_{n=1}^3{s_n r_n \over z(t) - p_n} \right| \right\}
$$
is minimal. For the four-point contact 
interaction, we have three finite poles and thus eight possible 
combinations of signs to try. Finally, we need the expression 
\ben
\hspace{-12pt} \int_{-1}^1{{\sqrt{1-t^2} \over z(t) - p_n} dt} 
&\hspace{-4pt}=\hspace{-4pt}& {2 \over z_2-z_1} 
\int_{-1}^1{{\sqrt{1-t^2} \over t + q} dt} \,, 
\qquad q = {z_1 + z_2 - 2 p_n \over z_2 - z_1} 
\nonumber \\ 
&\hspace{-4pt}=\hspace{-4pt}& {2 \pi \over z_2-z_1} \left( q - 
i \sqrt{1-q^2} \times \left\{ \begin{array}{ll} 
\sgn(\Im q) \ &{\rm if} \ \Im q \neq 0 \\
- \sgn(q) \ & {\rm if} \ \Im q = 0 \end{array} 
\right. \right) \,. \label{intsqrt}
\een
Now that we have subtracted the poles, the integral
\be
\int_{-1}^1{\left( 
{\csqrt{(z(t)-z_3) (z(t) - z_4)} \over z(t) (z(t)-1) (z(t)-\xi)}
- \sum_{n=1}^3{s_n r_n \over z(t) - p_n} \right) \sqrt{1-t^2} dt}
\label{sqrt_integrand}
\ee
can be computed very precisely\footnote{
To avoid cancellation errors (subtracting two quantities 
that are large and close to each other), we should explicitly cancel the
pole in (\ref{sqrt_integrand}). A simplified example is 
${\sqrt{t+1}\over t} - {1 \over t} = {1 \over \sqrt{t+1}+1} \,,$
and the pole at $t=0$ has been explicitly cancelled. A similar
simplification of the expression (\ref{sqrt_integrand}) is possible
but it gives complicated expressions. Instead of doing that, our program 
always chooses the two independent paths that are the furthest away 
from the poles, so that the cancellation errors are not too large.} 
with a Gauss method with weight
$\sqrt{1-t^2}$ (Jacobi weight function) (see for example \cite{Rals-Rabi}). 
We use the method of rank $30$ which, as we observe, has an 
accuracy of about twenty significant digits for 
$\xi = {1 \over 2} + {\sqrt{3} \over 2} i$.

%%%%%%%%%%%%%%%%%%%%%%%%%%%%%%%%%%%%%%%%%%%%%%%%%%%%%%%%%%%%%%%%%%%%%%%%%%%%%%
\subsection{Numerical evaluation of ${d \over da} \ell(z_i, z_j)$}
%%%%%%%%%%%%%%%%%%%%%%%%%%%%%%%%%%%%%%%%%%%%%%%%%%%%%%%%%%%%%%%%%%%%%%%%%%%%%%

Again, without loss of generality, we take $z_i = z_1$ and $z_j = z_2$.
We have
\ben
{d \over da} \ell(z_1, z_2) &=& 
{dz_2 \over da} \csqrt{\phi(z_2)} - {dz_1 \over da} \csqrt{\phi(z_1)} 
+ \int_{z_1}^{z_2}{{d \over da}\csqrt{\phi(z)} \, dz} 
\nonumber \\
&=& \int_{z_1}^{z_2}{{d \over da}\csqrt{\phi(z)} \, dz} 
\nonumber \\
&=& {1 \over 2} \int_{z_1}^{z_2}{{1 \over \csqrt{\phi(z)}} \,  
{1 \over z (z-1) (z-\xi)} \, dz}
\een
where we have used (\ref{phi1}), and the fact that $\phi(z)$ vanishes at 
$z_1$ and $z_2$. Now we use again the parametrization (\ref{z(t)}), and we find
\be
{d \over da} \ell(z_1, z_2) = -{1 \over 2} \int_1^{-1}{{1 \over 
\csqrt{(z(t)-z_3) (z(t) - z_4)}} {dt \over \sqrt{1-t^2}}} \,.
\label{derivative}
\ee
This integration\footnote{ 
We are not subtracting the poles here. Indeed, the accuracy that we
would gain would only make the Newton method slightly faster, without
effecting the accuracy of the final result.
}
can be done numerically with a Gauss method with weight 
$1/\sqrt{1-t^2}$. We use here the method of rank 30.

\begin{figure}[!ht]
\begin{center}
\input{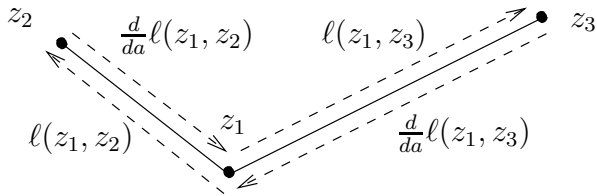}
\caption{\footnotesize{The arrows show the order in which the numerical integrations
should be done. In order for the signs of the continuous square root
to be consistent, the values used by the numerical integration methods 
should run continuously along the
closed path formed by the arrows, starting and ending at $z_1$.}}
\label{paths_f}
\end{center}
\end{figure}

It is important to note, as already mentioned after Equ.(\ref{int1}), 
that we have made the same choices of 
signs in both the calculations of $\ell(z_1, z_2)$ and 
${d \over da} \ell(z_1, z_2)$, namely
\be
\csqrt{\phi(z(t))} = {z_2-z_1 \over 2} \, {\csqrt{(z(t)-z_3) (z(t) - z_4)} 
\over z(t) (z(t)-1) (z(t)-\xi)} \, \sqrt{1-t^2} \,.
\ee
The relative signs of $\ell(z_1, z_2)$ and ${d \over da} \ell(z_1,
z_2)$ will therefore be consistent if the numerical integrations are
done one after the other, and if the path of the second integral is the 
reverse path of the first one (see Figure \ref{paths_f}). This is why we 
have written the limits of integration in (\ref{derivative}) in the 
opposite order as in (\ref{length}). In this way, the continuous square root 
will be evaluated at successive points that are close to each other, and we 
will thus have avoided any branch cut problem.

We observe that if the initial value of $a$ is not too far from the Strebel 
value, the Newton method converges to the Strebel solution. 

%%%%%%%%%%%%%%%%%%%%%%%%%%%%%%%%%%%%%%%%%%%%%%%%%%%%%%%%%%%%%%%%%%%%%%%%%%%%%%
\sectiono{Computation of the mapping radii}
\label{s_rho}
%%%%%%%%%%%%%%%%%%%%%%%%%%%%%%%%%%%%%%%%%%%%%%%%%%%%%%%%%%%%%%%%%%%%%%%%%%%%%%

%%%%%%%%%%%%%%%%%%%%%%%%%%%%%%%%%%%%%%%%%%%%%%%%%%%%%%%%%%%%%%%%%%%%%%%%%%%%%%
\subsection{The finite poles}
%%%%%%%%%%%%%%%%%%%%%%%%%%%%%%%%%%%%%%%%%%%%%%%%%%%%%%%%%%%%%%%%%%%%%%%%%%%%%%

The mapping radius at the puncture $p_n$ is given by (\cite{Belo-Zwie}):
\be
\ln \rho_n = \lim_{\epsilon \rightarrow 0} \left( \Im 
\int_{p_n+\epsilon}^{\tilde{z}}{\sqrt{\phi(z)} d z} + \ln \epsilon 
\right) \,, \label{rho_Bel_Zwi}
\ee
where $\tilde{z}$ is any point lying on the closest critical trajectory, $\epsilon$ 
is a positive real number, and the 
path of integration is chosen such that it avoids the branch cuts. Here we 
will take $\tilde{z}$ to be a zero, say $z_1$, that lies on the closest critical 
trajectory, and the path can be chosen to be a straight line if we replace 
the square root by the continuous square root. Moreover, since the residue of 
$\sqrt{\phi(z)}$ at $z=p_n$ is purely imaginary, (\ref{rho_Bel_Zwi}) doesn't 
change if we replace the lower integration limit by $p_n+\epsilon \alpha$, 
where $|\alpha|=1$. Thus
\be
\ln \rho_n = \lim_{\epsilon \rightarrow 0} \left( \Im 
\int_{p_n+\epsilon \alpha}^{z_1}{\csqrt{\phi(z)} dz} 
+ \ln \epsilon \right) \,, \qquad \alpha = {z_1-p_n \over |z_1-p_n|} \,,
\label{rho_lim}
\ee
and the sign of the continuous square root is chosen in such a way 
that the limit exists. The above expression is clearly very 
inconvenient for numerical evaluation. But we can remedy this by explicitly 
cancelling the singularity at $\epsilon = 0$. First, we note that
\be
\lim_{\epsilon \rightarrow 0}\left(\ln \epsilon - 
\left(-{1 \over \sqrt{z_1-p_n}} \int_{p_n+\epsilon \alpha}^{z_1}
{{\sqrt{z_1-z} \over z-p_n} \, dz} + 2 (\ln2 - 1) + \ln|z_1-p_n| 
\right) \right) = 0 \label{lneps} \,.
\ee
And thus
\ben
\ln \rho_n &=& \lim_{\epsilon \rightarrow 0} \left( \Im 
\int_{p_n+\epsilon \alpha}^{z_1}{\left( \csqrt{\phi(z)} - 
{i \over \sqrt{z_1-p_n}} {\sqrt{z_1-z} \over z-p_n} \right) dz} \right)
+ 2 (\ln2 -1) + \ln|z_1-p_n|  \nonumber \\
&=& \Im \int_{p_n}^{z_1}{\left( \csqrt{\phi(z)} - 
{i \over \sqrt{z_1-p_n}} {\sqrt{z_1-z} \over z-p_n} \right) dz}
+ 2 (\ln2 -1) + \ln|z_1-p_n| \,.
\een
Thus
\ben
\rho_n &\hspace{-4pt}=\hspace{-4pt}& {4 \over e^2} |z_1-p_n| \times 
\nonumber \\ 
&& \hspace{-12pt}\times \exp\left(\Im \left\{ (z_1-p_n) \int_0^1
{\left( {\csqrt{-(p_n-z_1)(z-z_2)(z-z_3)(z-z_4)} \over z (z-1)(z-\xi)} 
- {i \over z-p_n} \right) \sqrt{t} \, dt} \right\} \right) , 
\nonumber \\
\label{finite_pole}
\een
where we have parameterized $z = z_1 + t(p_n-z_1)$. The sign of the 
continuous square root is chosen so that the integrand is finite  
for $z \rightarrow p_n$. The integral is computed numerically with a 
rank 30 Gauss method with weight $\sqrt{t}$.

We haven't yet said how to know which zeros lie on the closest critical
trajectory. One method would be to actually compute the critical
trajectory, but this would require solving a differential equation, 
which can be hard numerically, and is certainly time consuming. Another method,
that we are using here\footnote{This particular method does not generalize 
to vertices of higher order. Indeed at order five we already have six zeros, 
and discarding three zeros out of them is thus impossible. For these higher 
vertices, we can use a fixed labelling of the zeros and poles and always make 
sure that the topology of the critical graph doesn't change, so that we always 
know which zeros are associated with which poles (see \cite{quintic}).
}, consists of calculating the radius for {\em all
four} zeros $z_i$. One of the four results may be different from
the other three, we discard it since it must correspond to the zero
not lying on the closest critical trajectory. Practically, we discard the 
value that is the furthest away from the average value.
The redundancy of calculations can then be used 
to evaluate the numerical errors affecting the
mapping radii. If we denote by $\rho_n^{(i)}$ the mapping radius for
the pole $p_n$ calculated with the zero $z_i$, and if $z_j$ is 
the discarded zero, we identify $\rho_n$ with the average
$\rho_n \approx \bar{\rho_n} = {1 \over 3} 
\sum_{i=1 \atop i \neq j}^4 \rho_n^{(i)}$. And we can estimate 
the variance of this sample $\sigma^2_{\rho_n} \approx {1 \over 2} 
\sum_{i=1 \atop i \neq j}^4 (\rho_n^{(i)} - \bar{\rho_n})^2$.

%%%%%%%%%%%%%%%%%%%%%%%%%%%%%%%%%%%%%%%%%%%%%%%%%%%%%%%%%%%%%%%%%%%%%%%%%%%%%%
\subsection{The pole at infinity}
%%%%%%%%%%%%%%%%%%%%%%%%%%%%%%%%%%%%%%%%%%%%%%%%%%%%%%%%%%%%%%%%%%%%%%%%%%%%%%

Here we must calculate the mapping radius in the coordinate 
$w=1/z$, in which the pole is at $w=0$. From $\phi(w) (dw)^2 = 
\phi(z) (dz)^2$ we have that $\phi(w) = \phi(z=1/w) / w^4$. Thus
\ben
\ln \rho_4 &=& \lim_{\epsilon \rightarrow 0} \left( \Im 
\int_{\epsilon \alpha}^{1/z_1}{\csqrt{\phi(w)} dw} 
+ \ln \epsilon \right) \,, \qquad \alpha = {z_1^*\over |z_1|} 
\nonumber \\
&=& \Im \int_{0}^{1/z_1}{\left({\csqrt{\phi(z=1/w)} \over w^2} 
- i {\sqrt{1-w z_1} \over w} \right) dw} + 2 (\ln 2-1) - \ln |z_1| \,,
\nonumber
\een
where $z_1$ is a zero that lies on the closest trajectory to the pole 
at infinity. With the parametrization $w = 1/z_1 - t/z_1$, we find
\be
\rho_4 = {4 \over e^2 |z_1|} 
\exp\left(\Im \left\{ {1 \over z_1} \int_0^1{\left( {\csqrt{-(1-w z_2) 
(1- w z_3) (1- w z_4)} \over w (1-w) (1 - w \xi)} - {i \over w} \right)
\sqrt{t} \, dt} \right\} \right) \,.
\ee
This expression is numerically calculated in the same way as (\ref{finite_pole}).

%%%%%%%%%%%%%%%%%%%%%%%%%%%%%%%%%%%%%%%%%%%%%%%%%%%%%%%%%%%%%%%%%%%%%%%%%%%%%%
\sectiono{The results as fits}
\label{s_results}
%%%%%%%%%%%%%%%%%%%%%%%%%%%%%%%%%%%%%%%%%%%%%%%%%%%%%%%%%%%%%%%%%%%%%%%%%%%%%%

The results obtained by our program are lengthy tables of data, namely the 
parameter $a$ and the mapping radii $\rho_n$ given at arbitrarily many points of 
the portion of moduli space $\A$. We do not write here any such table, but instead 
we express our results in the form of fits. Although some precision is lost 
with the fits, they are reasonably easy to use. The fits that we give below all 
have an error of the order of $0.1 \%$, which should be precise enough for many 
calculations.

%%%%%%%%%%%%%%%%%%%%%%%%%%%%%%%%%%%%%%%%%%%%%%%%%%%%%%%%%%%%%%%%%%%%%%%%%%%%%%
\subsection{The boundary of the moduli space}
%%%%%%%%%%%%%%%%%%%%%%%%%%%%%%%%%%%%%%%%%%%%%%%%%%%%%%%%%%%%%%%%%%%%%%%%%%%%%%

\begin{figure}[!ht]
\begin{center}
\input{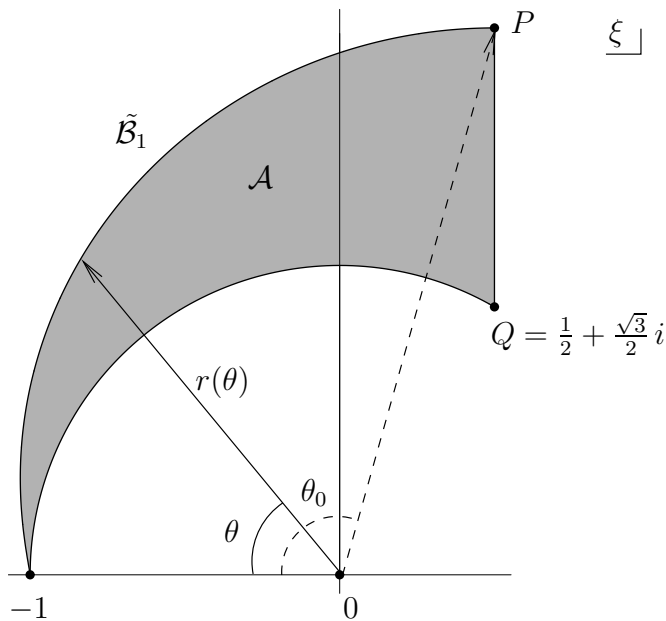}
\caption{\footnotesize{The parametrization of the curve $\tilde{\B_1}$ with the function 
$r(\theta)$.}}
\label{polar_f}
\end{center}
\end{figure}
The element $\A$ of moduli space that we are concentrating on, has 
$\Re \xi \leq 1/2$, $\Im \xi \geq 0$, and it is bounded by the unit circle 
centered at the origin, and by the curve $\B_1$ (see Figure \ref{moduli_f}). 
To describe $\A$, we 
thus need to describe the part of $\B_1$ with $\Re \xi \leq 1/2$ and 
$\Im \xi \geq 0$, which we denote by $\tilde{\B_1}$. We parameterize this curve 
in terms of the angle $\theta$ (see Figure \ref{polar_f}), with the expression 
\ben
\xi(\theta) = -r(\theta) \, e^{-i \theta} \,, 
\een
where $\theta \in [0,\, \theta_0]$, and $\theta_0 = 1.84644$ corresponds to 
the corner $P$ with $\Re \xi = 1/2$. We will use the following fit
\ben
\xi^{\rm fit}(\theta) = -r^{\rm fit}(\theta) \, e^{-i \theta} \,, 
\een
where $r^{\rm fit}(\theta)$ is a polynomial of order five in $\theta$
\be
r^{\rm fit}(\theta) = 1 + \sum_{i=1}^5{r_i \, \theta^i} \,. \label{fit1}
\ee
Note that the leading term has been set to one. In this way, the point $\xi = -1$, which 
we know is a corner of the moduli space (see Section \ref{s_moduli}) and belongs to 
$\tilde{\B_1}$, 
will also belong to the fit of $\tilde{\B_1}$ (at $\theta=0$); and the corner will 
be fitted properly.

The coefficients $r_i$ in (\ref{fit1}), are found by minimizing 
\be
\epsilon_r = \sqrt{\int_{\tilde{\B_1}}{\left( r^{\rm fit}(\theta) - r(\theta) \right)^2 ds} \over 
\int_{\tilde{\B_1}}{ds}} \,. \label{epsr}
\ee
Here $ds$ is the length element; the denominator in the square root of (\ref{epsr}) 
is thus the length of $\tilde{\B_1}$ and has been put there in order to normalize the error.
We find
\be
r^{\rm fit}(\theta) = 1 + 0.196639 \, \theta + 0.264670 \, \theta^2 - 0.130932 \, \theta^3 
+ 0.062563 \, \theta^4 - 0.015452 \, \theta^5 \,,
\ee
with error $\epsilon_r = 0.00022$. It is also useful to measure the maximum error 
made by this fit
\be
\delta_r = \max_{\theta \in [0,\theta_0]}\left\{\left|r^{\rm fit}(\theta) - r(\theta) \right|\right\} = 
0.00088 \,.
\ee

%%%%%%%%%%%%%%%%%%%%%%%%%%%%%%%%%%%%%%%%%%%%%%%%%%%%%%%%%%%%%%%%%%%%%%%%%%%%%%
\subsection{The Strebel differential}
%%%%%%%%%%%%%%%%%%%%%%%%%%%%%%%%%%%%%%%%%%%%%%%%%%%%%%%%%%%%%%%%%%%%%%%%%%%%%%

We write $\xi = x + y \, i$. And we use a polynomial fit of total order $5$ 
in $x$ and $y$
\be
a^{\rm fit}(\xi) = \sum_{i+j \leq 5}{a_{i j} \, x^i y^j} \,, \qquad \xi \in \A \,.
\ee
The coefficients $a_{i j}$ are found by minimizing the error
\be
\epsilon_a = \sqrt{\int_{\A}{\left| a^{\rm fit}(\xi) - a(\xi) \right|^2 dx dy} \over 
\int_{\A}{dx dy}} \,.
\ee
We find
\ben
a^{\rm fit}(\xi) &\hspace{-6pt}=\hspace{-6pt}& (0.67867 + 0.46345\,i) 
+ (2.07710 + 0.49176\,i) \,x + (0.20000 + 0.58477\,i) \,y \nonumber \\
&& \hspace{-8pt} + (1.61074 - 0.64246\,i) \,x^2 - (0.09194 - 0.36467\,i) \,x y 
+ (0.05965 + 0.38101\,i) \,y^2 \nonumber\\ 
&& \hspace{-8pt} - (0.33945 + 0.80519\,i) \,x^3 - (0.35448 - 0.94778\,i) \,x^2\,y 
- (1.04747 + 0.85263\,i) \,x y^2 \nonumber \\
&& \hspace{-8pt} + (0.56049 - 0.43233\,i) \,y^3 - (0.63061 + 0.05352\,i) \,x^4 
+ (0.27490 + 0.76364\,i) \,x^3 y \nonumber \\
&& \hspace{-8pt} - (0.89712 + 0.75355\,i) \,x^2 y^2 + (0.70167 + 0.46374\,i) \,x y^3 
- (0.50743 - 0.12128\,i) \,y^4 \nonumber \\
&& \hspace{-8pt} - (0.07990 - 0.08258\,i) \,x^5 + (0.32821 + 0.08618\,i) \,x^4 y 
- (0.08918 + 0.20755\,i) \,x^3 y^2 \nonumber \\
&& \hspace{-8pt} + (0.37843 + 0.20916\,i) \,x^2 y^3 - (0.15046 + 0.08495\,i) \,x y^4 
+ (0.11992 - 0.00719\,i) \,y^5 \nonumber \\
\een
with $\epsilon_a = 0.0011$. The maximum error made by this fit is
\be
\delta_a = \max_{\xi \in \A}\left\{ \left| a^{\rm fit}(\xi) - a(\xi) \right | \right\} = 0.0037 \,.
\ee

%%%%%%%%%%%%%%%%%%%%%%%%%%%%%%%%%%%%%%%%%%%%%%%%%%%%%%%%%%%%%%%%%%%%%%%%%%%%%%
\subsection{The mapping radii}
%%%%%%%%%%%%%%%%%%%%%%%%%%%%%%%%%%%%%%%%%%%%%%%%%%%%%%%%%%%%%%%%%%%%%%%%%%%%%%

We use again polynomials of order 5
\be
\rho_n^{\rm fit}(\xi) = \sum_{i+j \leq 5}{\rho_{n, i j} \, x^i y^j} \,,
\ee
but here we minimize the average of the square of the {\it relative} errors
\be
\epsilon_{\rho_n} = \sqrt{\int_{\A}{\left( {\rho_n^{\rm fit}(\xi) - \rho_n(\xi) \over \rho_n(\xi)} 
\right)^2 dx dy} \over 
\int_{\A}{dx dy}} \,. 
\ee
This is more natural than the absolute error, because the mapping radii appear as a 
fundamentally multiplicative factor (see \ref{conf2}); in some sense, 
they measure the sizes of the images of the unit disks.
The results of the fits are
\ben
\rho_1^{\rm fit}(\xi) &\hspace{-6pt}=\hspace{-6pt}& 0.31883 + 0.14609 \, x + 0.18216 \, y 
+ 0.47185 \, x^2 + 0.14573 x y + 0.15017 y^2 \nonumber \\
&& \hspace{-8pt} + 0.01342 \, x^3 - 0.11152 \,x^2 y - 0.29992 \,x y^2 + 0.02631\,y^3 
- 0.15920 \, x^4 + 0.01005 \,x^3 y \nonumber \\ 
&& \hspace{-8pt} - 0.28574 \,x^2 y^2 + 0.17921 \,x y^3 - 0.10879 \,y^4 - 0.02838 \, x^5 
+ 0.08245 \,x^4 y - 0.01874 \,x^3 y^2 \nonumber \\ 
&& \hspace{-8pt} + 0.12423 \,x^2 y^3 - 0.03863\,x y^4 + 0.03250\,y^5 \,,
\een
\ben
\rho_2^{\rm fit}(\xi) &\hspace{-6pt}=\hspace{-6pt}& 0.78098 - 0.59124\,x + 0.17253\,y 
- 0.40055\,x^2 + 0.44703\,x y - 0.10848\,y^2 \nonumber \\
&& \hspace{-8pt} + 0.10574\,x^3 + 0.27517\,x^2 y - 0.03518\,x y^2 - 0.05091\,y^3 
+ 0.15196\,x^4 - 0.08497\,x^3 y \nonumber \\
&& \hspace{-8pt} + 0.05597\,x^2 y^2 - 0.08168\,x y^3 + 0.07637\,y^4 + 0.01799\,x^5 
- 0.08381\,x^4 y + 0.02112\,x^3 y^2 \nonumber \\
&& \hspace{-8pt} - 0.05276\,x^2 y^3 + 0.02554\,x y^4 - 0.01999\,y^5 \,,
\een
\ben
\rho_3^{\rm fit}(\xi) &\hspace{-6pt}=\hspace{-6pt}& 0.29797 - 0.36138\,x + 0.18030\,y 
+ 0.17997\,x^2 + 0.49329\,x y + 0.32231\,y^2 \nonumber \\
&& \hspace{-8pt} - 0.33706\,x^3 + 0.61497\,x^2 y - 0.72821\,x y^2 + 0.23999\,y^3 
- 0.16258\,x^4 + 0.41692\,x^3 y \nonumber \\
&& \hspace{-8pt} - 0.71139\,x^2 y^2 + 0.46090\,x y^3 - 0.24944\,y^4 + 0.01369\,x^5 
+ 0.11153\,x^4 y - 0.13361\,x^3 y^2 \nonumber \\
&& \hspace{-8pt} + 0.21281\,x^2 y^3 - 0.10159\,x y^4 + 0.05898\,y^5 \,,
\een
\ben
\rho_4^{\rm fit}(\xi) &\hspace{-6pt}=\hspace{-6pt}& 0.95136 + 0.48365\,x - 0.36818\,y 
- 0.16100\,x^2 - 0.52786\,x y - 0.01609\,y^2 \nonumber \\
&& \hspace{-8pt} - 0.20922\,x^3 + 0.00970\,x^2 y + 0.19654\,x y^2 + 0.02470\,y^3 
- 0.00082\,x^4 + 0.17750\,x^3 y \nonumber \\
&& \hspace{-8pt} + 0.08937\,x^2 y^2 - 0.01417\,x y^3 + 0.01641\,y^4 
+ 0.01554\,x^5 + 0.00976\,x^4 y - 0.03639\,x^3 y^2 \nonumber \\
&& \hspace{-8pt}- 0.03248\,x^2 y^3 - 0.00411\,x y^4 - 0.00697\,y^5 \,.
\een
The errors are respectively
\be
\begin{array}{rcl}
\epsilon_{\rho_1} &=&  0.00051 \\
\epsilon_{\rho_2} &=&  0.00025 \\
\epsilon_{\rho_3} &=&  0.00035 \\
\epsilon_{\rho_4} &=&  0.00028 \,,
\end{array}
\ee
while the maximum relative errors $\delta_{\rho_n} = \max_{\xi \in \A}\left\{ 
\left| \rho_n^{\rm fit}(\xi) - \rho_n(\xi) \right| / \rho_n(\xi) \right\}$ are
\be
\begin{array}{rcl}
\delta_{\rho_1} &=& 0.0020 \\
\delta_{\rho_2} &=& 0.0009 \\
\delta_{\rho_3} &=& 0.0010 \\
\delta_{\rho_4} &=& 0.0014 \,.
\end{array}
\ee

%%%%%%%%%%%%%%%%%%%%%%%%%%%%%%%%%%%%%%%%%%%%%%%%%%%%%%%%%%%%%%%%%%%%%%%%%%%%%%
\sectiono{The four-tachyon contact term}
\label{s_4tachyon}
%%%%%%%%%%%%%%%%%%%%%%%%%%%%%%%%%%%%%%%%%%%%%%%%%%%%%%%%%%%%%%%%%%%%%%%%%%%%%%

As en elementary check of our results, we compare the easiest quantity that is 
calculable from our data, namely the four-tachyon contact term, with a result 
previously obtained, with a different method, in \cite{Belo}. 
It appears as the coefficient $v_4$ in the 
classical tachyonic potential 
\be
V(t) = -t^2 - \sum_{N=3}^{\infty}{v_N \, {t^N \over N!}}
\ee
(note that this is not the effective potential, it is the potential of the tachyon when 
we ignore all other fields). The coefficients $v_N$ are given by 
(\cite{Belo}, \cite{Belo-Zwie})
\be
v_N = (-1)^N {2 \over \pi^{N-3}} \int_{\V_{0,N}} \left( \prod_{n=1}^{N-3}
{d^2 \xi_n} \right) \prod_{n=1}^N{\rho_n^{-2}} \,,
\ee
where $\V_{0,N}$ is the restricted moduli space of the N-punctured sphere, and 
$\rho_n$ is the mapping radius at the puncture $\xi_n$. Thus
\ben
v_4 &=& {2 \over \pi} \int_{\V_{0,4}}(\rho_1 \rho_2 \rho_3 \rho_4)^{-2} \, 
d^2 \xi \nonumber \\
&=& {24 \over \pi} \int_{\A}(\rho_1 \rho_2 \rho_3 \rho_4)^{-2} \, d^2 \xi \,,
\een
where we have used the fact that the measure $(\rho_1 \rho_2 \rho_3 \rho_4)^{-2} \, d^2 \xi$, 
is invariant under the action of ${\rm PSL}(2, \mathbb{C})$ transformations that permute the 
fixed poles $\{0, 1, \infty\}$, and under complex conjugation; the integration over 
$\V_{0,4}$ can thus be replaced by 12 times the integration over the twelfth of moduli space 
$\A$. Using our full data, we get
\be
v_4 = 72.390 \pm 0.003 \,, \label{v4}
\ee
in agreement with the result $v_4 = 72.39$ found in \cite{Belo}. The uncertainty 
in (\ref{v4}) was computed from the uncertainties in the mapping radii 
(see Section \ref{s_rho}), and the uncertainty from the numerical integration 
over $\A$, which was done with a Monte-Carlo technique.

It is also interesting to compare this to the same calculation done
with the fits given in Section \ref{s_results}. We obtain $v^{\rm
fit}_4 = 72.41$, less than $0.03 \%$ away from the value (\ref{v4}).

%%%%%%%%%%%%%%%%%%%%%%%%%%%%%%%%%%%%%%%%%%%%%%%%%%%%%%%%%%%%%%%%%%%%%%%%%%%%%%
\sectiono{Conclusions}
\label{s_conclusions}
%%%%%%%%%%%%%%%%%%%%%%%%%%%%%%%%%%%%%%%%%%%%%%%%%%%%%%%%%%%%%%%%%%%%%%%%%%%%%%

We have described numerically the geometry of the contact interaction
of four off-shell closed bosonic string states. Our results give the
Strebel differentials and the mapping radii everywhere in the twelfth
of moduli space $\A$, from which we can extend the results to the
whole moduli space $\V_{0,4}$ by straightforward ${\rm PSL}(2,
\mathbb{C})$ transformations and complex conjugations. We have
illustrated how the knowledge of the Strebel differentials and the
mapping radii can be used to calculate the local coordinates around
the punctures of the Riemann sphere. Our results can therefore be used
to insert off-shell vertex operators on the four-punctured sphere,
which is needed to compute off-shell contact amplitudes of four closed
string states.

\paragraph{}
The numerical methods that we have developed can be generalized to
contact interactions of more than four strings. Actually, the 
motivation for the present work was to compute the bulk tachyon
potential and investigate whether it has a minimum or not. While the
present paper doesn't offer any new information about the tachyon
potential, it confirms its quartic term, previously calculated in
\cite{Belo} with a different method. A generalization of the work
presented here, to quintic order in the string field, is in progress
(\cite{quintic}) and will hopefully give the tachyon potential to
order five.

\paragraph{}
It would be interesting to use the results of this paper for doing 
concrete calculations in CSFT to quartic order in diverse backgrounds, 
for example in 
an orbifold background (\cite{orbifolds, Okaw-Zwie1}). Another 
interesting application would be to do computations in heterotic string 
field theory (\cite{Okaw-Zwie2}).

\section*{Acknowledgments}
I am indebted to Barton Zwiebach for many useful discussions. I also
thank Yoji Michishita, Martin Schnabl and Peter West for useful
conversations. And I would like to thank Jean-Pierre Derendinger and
the Institute of Physics of the University of Neuch\^atel, where part
of this work was done, for their hospitality during the summer of
2003. This work has been supported in part by the Swiss National
Science Foundation, and by the PPARC rolling grant:
PPA/G/O/2002/00475.

%%%%%%%%%%%%%%%%%%%%%%%%%%%%%%%%%%%%%%%%%%%%%%%%%%%%%%%%%%%%%%%%%%%%%%%%%%%%%%
%\newpage

\end{document}